\documentclass[prl,twocolumn,superscriptaddress,showpacs,a4paper]{revtex4}
\usepackage{booktabs}
\usepackage{pdflscape}
\usepackage{bm}        
\usepackage{amssymb} 
\usepackage{amsmath}
\usepackage{color}
\usepackage{tensor}
\usepackage{manfnt}

\begin{document}


\title{$\mathbb{Z}_2$ Lattice Gerbe Theory}

\date{1st June 2014}

\author{Desmond A.\ Johnston}
\affiliation{Department of Mathematics, 
School of Mathematical and Computer Sciences,
Heriot-Watt University, Riccarton, Edinburgh EH14 4AS, Scotland}

\begin{abstract}
Discretized formulations of 
$2$-form abelian \cite{savit, pearson, orland, froh, omero, baig, halliday} and non-abelian \cite{nepomechie1, orland2, orland3, orland4} gauge fields on $d$-dimensional hypercubic lattices have been discussed 
in the past by various authors and most recently in \cite{1}.
In this  note we recall that the  Hamiltonian of a $\mathbb{Z}_2$
variant of such theories is one of the family of generalized Ising models originally considered by Wegner  \cite{2}. For such ``$\mathbb{Z}_2$ lattice gerbe theories'' general arguments  can be used to show that a phase transition for Wilson surfaces will occur for $d>3$ between volume and area scaling behaviour. In $3d$ the model is equivalent under duality to an  infinite coupling model and no transition is seen, whereas in $4d$ the model is dual to the  $4d$ Ising model and displays a continuous transition.  In $5d$ 
the   $\mathbb{Z}_2$ lattice gerbe theory is self-dual in the presence of an external field and in $6d$ it is self-dual in zero external field.

\end{abstract}


 
\maketitle


The study of the lattice discretization of $1$-form gauge fields, lattice gauge theory,  is now a mature subject. Higher form gauge fields on the lattice have received less attention, although $2$-form gauge fields or gerbes have also proved to be of considerable interest. In string theory interest in higher form fields dates back to the work of Kalb and Ramond \cite{kr}  and they have also arisen more recently in M-theory, particularly with regard to describing M5-branes. Abelian gerbes suffice to describe the six-dimensional low energy effective theory arising from a single M5-brane \cite{s3,s4}, but multiple M5-branes  require a non-abelian theory. 
This has proved difficult to formulate in the continuum and motivated the authors of \cite{1} to revisit both abelian and non-abelian gerbe theories on Euclidean hypercubic lattices in the manner of the Wilsonian formulation of lattice gauge theory \cite{3,4,5,elitzur,creutz2,ukawa}. The theory of  abelian $2$-form (and also higher $n$-form) fields on a lattice had already been explored from varying perspectives  in \cite{savit, pearson, orland, froh, omero, baig, halliday} and non-abelian lattice theories were also proposed in \cite{nepomechie1, orland2, orland3, orland4}.
Later, rather similar ideas were pursued in \cite{larson, wise, lundow}. 
In the lattice gerbe theories the gauge variables lived on the faces of the cubes of a hypercubic lattice and the Hamiltonian (i.e. action in field theoretic language) was given by the product of face variables round elementary cubes, which were called cubets in \cite{1}, summed over the lattice.

We take a step further back here and consider what might be learnt from the simplest Hamiltonians that maintain the $2$-form gauge invariance, which are constructed from $\mathbb{Z}_2$ spins and which we call $\mathbb{Z}_2$ lattice gerbe theories, borrowing the nomenclature of \cite{1}.
Many of the characteristic features of a lattice gauge theory are already discernible in the simplest $\mathbb{Z}_2$ variant, so
a reasonable expectation is that such a $\mathbb{Z}_2$ lattice gerbe theory might offer a similar degree of insight into the behaviour of more general, including non-abelian, lattice gerbe theories. 

In this paper we note that the  Hamiltonians of the $\mathbb{Z}_2$
variant of  lattice gerbe theories are amongst those considered by Wegner \cite{2} (see also \cite{savit2}) and use this to discuss the phase structure of  the $\mathbb{Z}_2$ lattice gerbe theories in various dimensions, both with and without external fields.

It is possible to cast the Hamiltonians of abelian $n$-form theories on a hypercubic lattice for arbitrary $n$ in a common notation \cite{omero} by denoting the 
oriented $n$-dimensional cells of the $d$-dimensional hypercubic lattice by $C_n$ and
the cells with reversed orientation 
by $-C_n$. $n$-forms $A$ are then maps from the $C_n$ to $\mathbb{R},\, \mathbb{Z}, \, U(1), \ldots$ which satisfy $A(C_n) = - A(-C_n)$.
The Hamiltonians may then be written in the $U(1)$ case as
\begin{equation}
H=  -  \sum_{C_{n+1}} \left( \prod_{C_n \in \partial C_{n+1}} U(C_n)  + c.c. \right)
\label{ngauge}
\end{equation}
where the gauge variables $U (C_n)= \exp ( i A (C_n))$ live on the boundaries $C_n$ of cells $C_{n+1}$ and $c.c.$ denotes a complex conjugate. 
The Hamiltonian is thus given by the sum of products of the $U(C_n)$ around the boundary of a $C_{n+1}$.  

Specializing to a lattice gauge theory ($1$-form) on  a hypercubic lattice, the $U (C_1)$ live on edges, $C_1$, and the Hamiltonian is a sum of terms composed of products of $4$ such edge variables around the boundaries of two-dimensional faces or plaquettes, $C_2$. Similarly, for a lattice gerbe theory ($2$-form) the $U (C_2)$ live on faces, $C_2$, and the Hamiltonian is a sum of terms composed of products of $6$ such face variables around the boundaries of three-dimensional cubes, or cubets, $C_3$ as in \cite{1}. 

In lattice $n$-form theories the gauge transformations are implemented as 
\begin{equation}
U(C_n) \to \prod_{C_{n-1} \in \partial C_n} \Lambda (C_{n-1}) U(C_n) \, ,
\end{equation}
so in the  lattice gauge theory the gauge transformations $\Lambda(C_0)$ operate on the two sites at the end of each edge and in the lattice gerbe theory the $\Lambda (C_1)$ operate on the four edges round each face. 

The lattice gauge theory and $n$-form generalizations have no local order parameters, but can still display distinct confined and deconfined phases characterized by the behaviour of Wilson loops/surfaces \cite{2,3,4,5,elitzur,creutz2}. In the case of the gauge theory the different phases can be exposed by defining
\begin{equation}
\Gamma (L) = \left\langle   \prod_{{C_1} \in L} U (C_1)  \right\rangle 
\end{equation}
where $L$ is some closed loop in the lattice formed by the product of the $C_1$
and the averages are taken with respect to the Boltzmann weights $\exp ( - \beta H)$.
$\Gamma(L)$ would then be expected to scale as 
\begin{equation}
\Gamma (L) \sim 
\begin{cases} 
\exp ( - A(L) ) & \beta < \beta_c \\ 
\exp ( - P(L) )  &  \beta > \beta_c
\end{cases}
\label{area_perim}
\end{equation}
where $A(L), P(L)$ are the area and perimeter of the loop in a theory with a confining transition at $\beta_c$. 
An example of such behaviour is the $3d$ $\mathbb{Z}_2$ gauge theory, which is dual to the $3d$ Ising model and hence displays a
continuous phase transition. 

Similarly,  Wilson surface observables in the lattice gerbe theory may be defined as 
\begin{equation}
\Gamma (S) = \left\langle   \prod_{C_2 \in S} U (C_2 ) \right\rangle 
\end{equation}
where $S$ is some closed surface on the lattice formed by the product of the $C_2$. 
In a lattice gerbe theory with a deconfining transition  $\Gamma(S)$ would be expected to scale as, analogously to the lattice gauge theory,
\begin{equation}
\Gamma (S) \sim 
\begin{cases} 
\exp ( - V(S) ) & \beta < \beta_c \\ 
\exp ( - A(S) )  &  \beta > \beta_c
\end{cases}
\label{vol_area}
\end{equation} 
where $V(S), A(S)$ are the volume enclosed by and the surface area of the Wilson surface.
The authors of \cite{1} conducted some numerical investigations of  abelian lattice gerbe theory  and found volume scaling at all couplings in $3$ dimensions, whereas there was evidence of a phase transition from an  area law at weak coupling (large $\beta$) to a volume law at strong coupling (small $\beta$) in $6$ dimensions. 

In \cite{2} Wegner defined a class of generalized Ising models $M_{d,n}$ with $\mathbb{Z}_2$ spins living on $d$ dimensional hypercubic lattices which were characterized by a Hamiltonian of precisely the form specified in equ.~(\ref{ngauge}) with $M_{d,n}$ possessing $(n-1)$-form gauge symmetries. Since the $U(C_n)$
are now $\mathbb{Z}_2$ spins the complex conjugate is superfluous in this case.
In particular, a $\mathbb{Z}_2$ lattice gerbe theory on a $d$-dimensional hypercubic lattice is just the $M_{d,3}$ model considered by Wegner. It is  thus possible to use the general results of \cite{2} on duality and the scaling of Wilson loops and surfaces to make some observations on the phase structure
of $\mathbb{Z}_2$ lattice gerbe theories in various dimensions.

In general, on  a finite lattice composed of $N$ $d$-dimensional hypercubes, the $M_{d,n}$ model consists of $N_s = {d \choose n -1} N$ spins sited at the centres of the $(n-1)$-dimensional hypercubes and the Hamiltonians $H_{dn}$ consist of the product of $2 n$ spins on the $(n-1)$-dimensional faces of the $N_b = {d \choose n} N $ $n$-dimensional hypercubes. 
Duality arguments can be used to show that the partition functions of the models $M_{d,n}$ on the original $d$-dimensional hypercubic lattice and $M^*_{d, d-n}$ on the dual lattice (also a $d$-dimensional hypercubic lattice) are related for periodic boundary conditions by   
\begin{equation}
Z_{d,n} ( \beta) \sim Z^*_{d,d-n} ( \beta^*) \; ,
\label{dual}
\end{equation}
where inessential symmetrizing factors in the  relation between the partition functions
have been dropped and the couplings are related by
\begin{equation}
\tanh ( \beta) = \exp ( - 2 \beta^*) \, .
\label{couplingb}
\end{equation}
An external field term coupling the spins to a field $h$ can also be included in the standard manner in $M_{d,n}$. In this case the partition functions of the theories $M_{d,n}$ and $M^*_{d, d-n+1}$ are related by the duality 
\begin{equation}
Z_{d,n} ( \beta, h) \sim Z^*_{d,d-n+1} ( \beta^*, h^*) \; ,
\label{dualh}
\end{equation} 
where the  symmetrizing factors 
have again been dropped and the couplings $\beta, h$ in $M_{d,n}$ and $\beta^*, h^*$ in the dual theory were now related by
\begin{eqnarray}
\tanh ( \beta) &=& \exp ( - 2 h^*) \nonumber \\
\tanh  (h) &=& \exp ( - 2 \beta^*) \; .
\label{couplingsbh}
\end{eqnarray}
If the discussion is now restricted to the case $n=3$ relevant for the consideration of the lattice gerbe theory, the Hamiltonian contains the  product of $6$ face spins round a cubet, as in \cite{1}. 
Taking $d=3$ first, when $h=0$ equ.~(\ref{dualh})
reads 
\begin{equation}
Z_{3,3} ( \beta, 0) \sim Z^*_{3,1} ( \infty , h^*)
\label{dual33}
\end{equation}
where $\tanh (h^*) = \exp ( - 2 \beta)$. Since only the $\pm 1$ ground state spin configurations
contribute as $\beta^* \to \infty$ in the dual partition function $Z^*_{3,1} ( \beta^* , h^*)$ it is given in this limit by
\begin{equation}
Z^*_{3,1} (\beta^*, h^*)  \sim \exp(N_b \beta^* + N^*_s h^*) + \exp(N_b \beta^* - N^*_s h^*)
\end{equation}
where $N_b$ is the number of bonds connecting interacting spins. 
This can be translated back to the original, undualized partition function to give
\begin{equation}
Z_{3,3} (\beta, 0)  \sim 2^{N_s} \left[ \cosh(\beta)^{N_b} +\sinh(\beta)^{N_b} \right]
\end{equation}
which is analytic in $\beta$, so there is no transition for a $\mathbb{Z}_2$ lattice gerbe theory in $3d$.

For $d>3$ the situation is different. Low and high-temperature expansions \cite{2} reveal precisely the transition from area to volume scaling as $\beta$ is decreased that is indicated by equ.~(\ref{vol_area}). This confirms that a confining phase transition exists so long as $d>3$. It is also possible to infer that the critical coupling $\beta_c$ is a decreasing function of $d$, i.e. $\beta_{c, d+1, n} \le \beta_{c, d, n}$. Interestingly, abelian $U(1)$ $n$-form theories are expected to have a phase transition only in $d \ge n+3$ dimensions \cite{froh}, so a phase transition will only appear for $d \ge 5$ dimensions for the abelian $2$-form theory rather than $d \ge 4$ in the $\mathbb{Z}_2$ case, a fact already noted in \cite{orland, omero}. The absence of a phase transition in $3d$ in the $U(1)$ model in the numerical simulations of \cite{1} is consistent with these and other earlier studies \cite{quevedo, polyakov, diamantini}.

It is possible to make a more explicit statement about the nature of the $\mathbb{Z}_2$ lattice gerbe theory  phase transition in $d=4$ using the zero field duality since 
\begin{equation}
Z_{4,3} ( \beta) \sim Z^*_{4,1} ( \beta^*)
\label{dual43} 
\end{equation}
and the $4d$ $\mathbb{Z}_2$ lattice gerbe theory is thus dual to the $4d$ Ising model and will display a continuous phase transition. Since $d=4$ is the upper critical dimension of the Ising model the exponents will be mean field (though there is still some discussion about the nature of the specific heat divergence, if it exists,  \cite{lundow}). The $M_{4,3}$ model has been simulated in \cite{baig} to compare with the 
predictions from duality from the $4d$ Ising model and in \cite{halliday} with a view to formulating a viable cluster algorithm.

Self-dual $\mathbb{Z}_2$ lattice gerbe theories exist both in the presence and absence of an external field $h$.
Using equ.~(\ref{dual}) and $\tanh ( \beta) = \exp ( - 2 \beta^*)$ from equ.~(\ref{couplingb})
\begin{equation}
Z_{2n, n} ( \beta) \sim Z^*_{2n ,n} ( \beta^*) \; ,
\label{selfdual2n}
\end{equation}
so the $\mathbb{Z}_2$ lattice gerbe theory on a $6d$ hypercubic lattice will be self-dual.
Similarly, employing equ.~(\ref{dualh}) shows
that
\begin{equation}
Z_{2n -1, n} ( \beta, h) \sim Z^*_{2n -1,n} ( \beta^*, h^*) \; ,
\label{selfdual}
\end{equation}
where the duality relation for the couplings is now that in equ.~(\ref{couplingsbh}).
A  $\mathbb{Z}_2$ lattice gerbe theory   on a $5d$ hypercubic lattice in an external field will thus also be self dual with the duality relations: $\tanh ( \beta) = \exp ( - 2 h^*), \,
\tanh  (h) = \exp ( - 2 \beta^*)$. From a numerical point of view the interest of self-duality often lies in 
pinning down the critical coupling $\beta_c$ using, for instance,  equ.~(\ref{couplingb}) in the zero field case, which may be written more symmetrically as
\begin{equation}
\sinh ( 2 \beta_{c,d,n}  ) \sinh( 2 \beta^*_{c,d,d-n} ) = 1 \; .
\end{equation} 
For the self-dual case in the absence of an external field ($d=2n$) this give $\beta_{c,2n,n} = \beta^*_{c,2n,n} = \frac{1}{2} \ln ( 1 + \sqrt{2})$.
The critical coupling of the $\mathbb{Z}_2$  gerbe theory ($n=3$) on a $6d$ hypercubic lattice is thus identical
to that of a nearest neighbour $2d$ Ising model ($n=1$) on the square lattice or a $4d$ Ising gauge theory ($n=2$) on a $4d$ hypercubic lattice.

The general considerations of \cite{2} do not, of course, say anything about the nature of any phase transitions which are present, unless they can be deduced from the dual theory as in equ.~(\ref{dual43}). The interest from the string/M-theoretic point of view of lattice gerbe theories lies in the possibility of defining a non-trivial continuum theory at a continuous phase transition point, particularly in the non-abelian case. From this perspective, a first order deconfining transition between the volume law and area law behaviour, such as that suggested by the numerical results in \cite{1} for the $6d$ $U(1)$ abelian theory, would not be of interest, whereas the continuous transition seen in $4d$ for the $\mathbb{Z}_2$  theory would.

There has not, to our knowledge, been any recent systematic numerical investigation of the nature of the lattice gerbe  transitions apart from the exploratory results presented in \cite{1} and the $4d$ results of \cite{baig,halliday}. Earlier numerical work on such systems is now of considerable vintage \cite{pearson}.
In this context it is likely that the further numerical exploration of 
$\mathbb{Z}_2$ lattice gerbe theories can offer useful insight into the behaviour of  physically relevant abelian and non-abelian lattice gerbe theories in the same manner that
$\mathbb{Z}_2$ lattice gauge theory has cast light on the general properties of lattice gauge theories. An obvious generalization is to consider the behaviour of $\mathbb{Z}_N$ spins in the lattice gerbe action
for $N>2$ in the expectation that the behaviour would approach that of the $U(1)$ model as $N \to \infty$, a program pursued in the early days of lattice gauge theories simulations \cite{elitzur, creutz2, ukawa}. As discussed in \cite{omero}, it is expected that the lower critical dimension of $\mathbb{Z}_N$ gerbe models will also be $4$ and also  that  they will display (at least) three distinct phases in $6$ dimensions for sufficiently large $N$ \cite{orland}.

Viewed purely as lattice spin systems the lattice gerbe theories offer an interesting laboratory for exploring degeneracy, gauge invariance, lattice topological defects and features of scaling at both continuous and first order phase transitions in various dimensions. 
The theories discussed here and in \cite{1} are of ``pure'' lattice gerbe theories. Coupling $n$-form $A$ and $(n-1)$-form $B$  abelian $U(1)$ gauge fields, in the manner of \cite{omero,sjrey}, 
\begin{eqnarray}
H &=&  -  \sum_{C_{n+1}} \left( \prod_{C_n \in \partial C_{n+1}} U(C_n)  + c.c. \right) \nonumber\\
&-& \lambda \sum_{C_{n}} \left( U^q (C_n) \prod_{C_{n-1} \in \partial C_{n}} \Phi (C_{n-1})  + c.c. \right) \,  ,
\label{ngauge2}
\end{eqnarray} 
where $q$ is the charge associated with an $(n-1)$-form $B( C_{n-1})$ and $\Phi ( C_{n-1}) = \exp ( i q B( C_{n-1}) )$, 
would be another obvious extension. These mixed Hamiltonians would  present a richer phase structure than that of the pure lattice gerbe theory, with the possibility of Higgs-type phases \cite{sjrey} in the manner of the original $\mathbb{Z}_2$
lattice gauge-Higgs theory \cite{frad_shenker,jong}. The  Hamiltonian for this is given (schematically) by
\begin{eqnarray}
H &=& - \sum_{\Box} U^4 - \lambda \sum_{\langle ij \rangle} U \sigma^2
\end{eqnarray}
where  the sums are over faces, $\Box$, and edges, $\langle ij \rangle$, for the edge gauge spins $U$ and site matter spins $\sigma$.
The $\mathbb{Z}_2$ lattice gerbe equivalent of this would be 
\begin{eqnarray}
H &=& - \sum_{\mbox{\mancube}} U^6 - \lambda \sum_{\Box}  U \sigma^4  
\end{eqnarray}
where  the sums are now over cubets, $\mbox{\mancube}$, and faces, $\Box$, with the $2$-form gauge  spins $U$ living on faces and the $1$-form gauge  spins $\sigma$ living on edges. 
The mean field calculation of \cite{sjrey} in the $U(1)$ case suggests that Higgs, Coulomb and confined phases are all present.  It would be interesting to explore the phase structure numerically in various dimensions for both the discrete and continuous abelian models in this case too.

A final comment, also remarked on in \cite{1}, is that just as for a lattice gauge theory gauge-fixing is not obligatory in a lattice gerbe theory.
  In the lattice gauge theory case a gauge may nonetheless be fixed by, for instance, fixing gauge variables on a maximal tree \cite{creutz_gauge} in order to permit the evaluation of gauge-variant quantities such as correlators or for numerical reasons. The reasoning which permits this, namely that no gauge invariant loops are fixed by this procedure, also applies in the lattice gerbe case if surfaces are substituted for links. It is therefore possible to fix the plaquette gauge variables on a spanning open surface on the lattice in the lattice gerbe theory case.  


\end{document}